\documentclass[showpacs,showkeys,prl,amsmath,amssymb,floatfix]{revtex4}
\usepackage{graphicx}% Include figure files
\usepackage{dcolumn}% Align table columns on decimal point
\usepackage{bm}% bold math
\usepackage{natbib}
\usepackage{epstopdf}
\usepackage{natbib}
\usepackage{amsmath} 
\usepackage[abs]{overpic}
\usepackage{subfigure}
\usepackage{color} 
\usepackage{tabularx}

\begin{document}
\title{Fluctuation assisted spreading of a fluid filled elastic blister}

\author{Andreas Carlson}
% \email{ }
%\author{L. Mahadevan$^{1,2}$}
%  
%\author{ }
%
\affiliation{Department of Mathematics, University of Oslo.}
%\affiliation{$^2$Kavli Institute for Bionano Science and Technology, Departments of Physics, and Organismic and Evolutionary Biology, Harvard University, Cambridge, USA.}
\date{\today}

\begin{abstract}
In this theoretical and numerical study, we show how spatially extended fluctuations can influence and dominate the dynamics of a fluid filled elastic blister as is deforms onto a pre-wetted solid substrate. To describe the blister dynamics, we develop a stochastic elastohydrodynamic framework that couples the viscous flow, the elastic bending of the interface and the noise from the environment. We deploy a scaling analysis to find the elastohydrodynamic spreading law $\hat R\sim \hat t^{1/11}$ a direct analogue to the capillary spreading of drops, while the inclusion of noise in our model highlights that the dynamics speed-up significantly  $\hat R\sim \hat t^{1/6}$ as local changes in curvature enhance the peeling of the elastic interface from the substrate. Moreover, our analysis identifies a distinct criterion for the transition between the deterministic and stochastic spreading regime, which is further illustrated by numerical simulations. 
\end{abstract}

\maketitle
The motion of an elastic interface near a solid substrate manifest itself in microelectronic engineering processes \cite{Chaudhury2005,Chaudhury2013EPL}, membrane adhesives \cite{Leong, Hosoi:2004}, synthetic lipid membrane systems \cite{Reister2008,Cantat:1998}, geology \cite{michaut,Lister2013PRL,Nesic}, as well as in a myriad of biological processes including adhesion of epithelial cell sheets, bio-films \cite{Wilking}, membrane blebbing \cite{Alert:2016} and cell adhesion \cite{Bell:1984,Giannone:2007}. The spatiotemporal dynamics of these interfaces is composed of the complex interplay between the deterministic forces associated with the interface dynamics i.e. viscous film flow, membrane bending and tension, and stochastic effects from the environment. %Theoretical tools are needed to describe these stochastic elastohydrodynamic flows at sub-micron length scales, to help classify the principal mechanisms that dominate the interface dynamics.

%At nanoscopic length scales, molecular effects start to unveil themselves and a continuum description of the fluid flow becomes questionable. Molecular dynamics simulations \cite{Moseler:2000,Johansson2015} simulations have been effective in demonstrating how molecular motion cause a deviation from deterministic hydrodynamic theory, suggesting that additional stochastic effects need to be included in the analysis. Interfacial flows at these length scales have a large surface-to-area ratio and are conspicuous examples of noise effects in spatially extended systems. In narrow geometries such as liquid films, the height of the film separating the solid substrate from the elastic interface can be $\sim 10$ nm, but extend laterally $\sim 10$ $\mu$m, thus making a molecular description practically unattainable where mean-field theoretical models \cite{Diez2016b} can describe the interplay between physical forces generated by membrane bending, the viscous flow and the spatially extended fluctuations from the environment \cite{grun2006,Davidovitch:2005}. 

Coarse-grained models and the inclusion of stochastic stress terms into a continuum formulation are alternative approaches to a molecular description \cite{Moseler:2000,Johansson2015,Diez2016b,Diez2016a,grun2006} to describe interface dynamics. The development of phenomenological hydrodynamic models with a thermal noise terms have a long history\cite{Landau1959,Uhlenbeck:1970} .% and dates back to the seminal work by Landau \cite{Landau1959} and further developed by Uhlenbeck and Fox \cite{Uhlenbeck:1970}.%
Capillary flows e.g. breakup of a liquid tread \cite{Moseler:2000}, rupture and de-wetting of thin liquid films \cite{Seemann2001,Herminghaus}, drop spreading \cite{Davidovitch:2005} and contact angle dynamics \cite{Belardinelli2016}, have been shown to critically depend on fluctuations at nanoscopic scale and to favorably compare with a stochastic hydrodynamic description. A canonical example of capillary flow is the spreading of a viscous droplet on a solid substrate, where the spreading radius $\hat R$ adopts an asymptotic solution in time $\hat t$, also known as Tanner's law i.e. $\hat R\sim \hat t^{{1}/{10}}$ and $\hat R\sim \hat t^{{1}/{7}}$ in one dimensional geometries. Lubrication theory gives an accurate description of Tanner's law \cite{tanner} and has been extended with stochastic effects from thermal fluctuations. Asymptotic analysis and numerical simulations show that thermal fluctuations can significantly alter the dynamics and accelerate the spreading rate $\hat R\sim \hat t^{{1}/{6}}$ and $\hat R\sim \hat t^{{1}/{4}}$  in one dimensional geometries \cite{Davidovitch:2005,Johansson2015}. Far less is known about how fluctuations alter elastohydrodynamic flows, although they are as generic \cite{Arutkin2017,Tulchinsky2017,Al-Housseiny:2013,Hewitt:2015,Pihler:2012} and play an intrinsic role in a myriad of biological membrane phenomena with a bending rigidity \cite{carlson:2015,Monzel2015,legoff2015}. Studies so far have focused on linear and non-conservative films in one dimension \cite{legoff2014,legoff2015}, which can overlook these non-linear interactions.

\begin{figure}[ht!]
\centering
\includegraphics[width=1.01\linewidth]{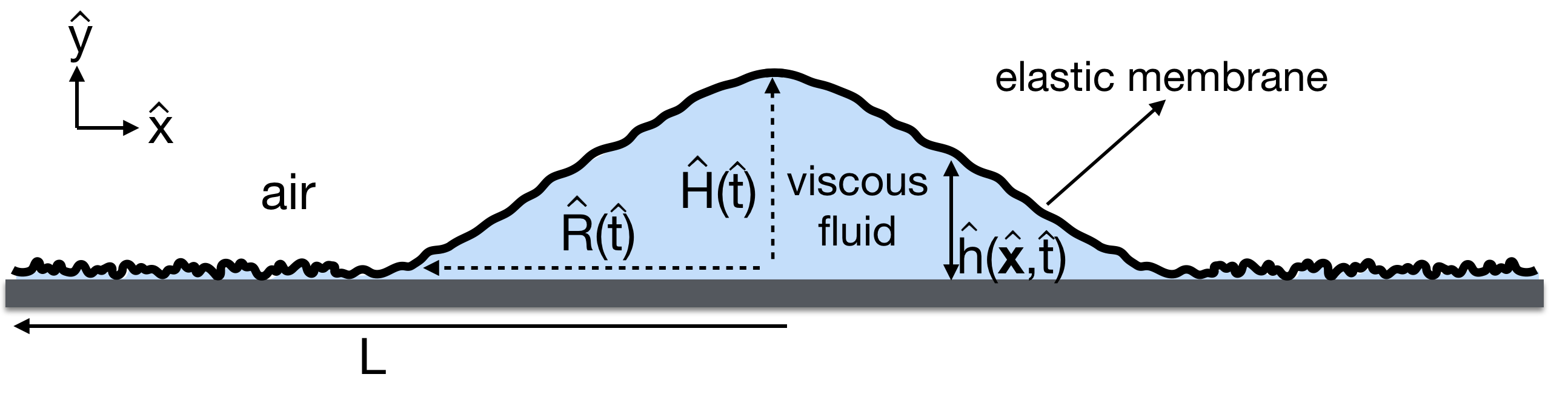}
\caption{A schematic of the fluid filled blister that has an elastic interface with a bending rigidity, placed on on a pre-wetted solid substrate with a microscopic film of height $\hat\epsilon$. The blister is described by its height field $\hat h(\mathbf{\hat x},\hat t)$ and has a characteristic height $\hat H(\hat t)$ and spreading radius $\hat R(\hat t)$. 
\label{fig:fig1}}
\end{figure}
\begin{figure}[h!]
\centering
\includegraphics[width=1.0\linewidth]{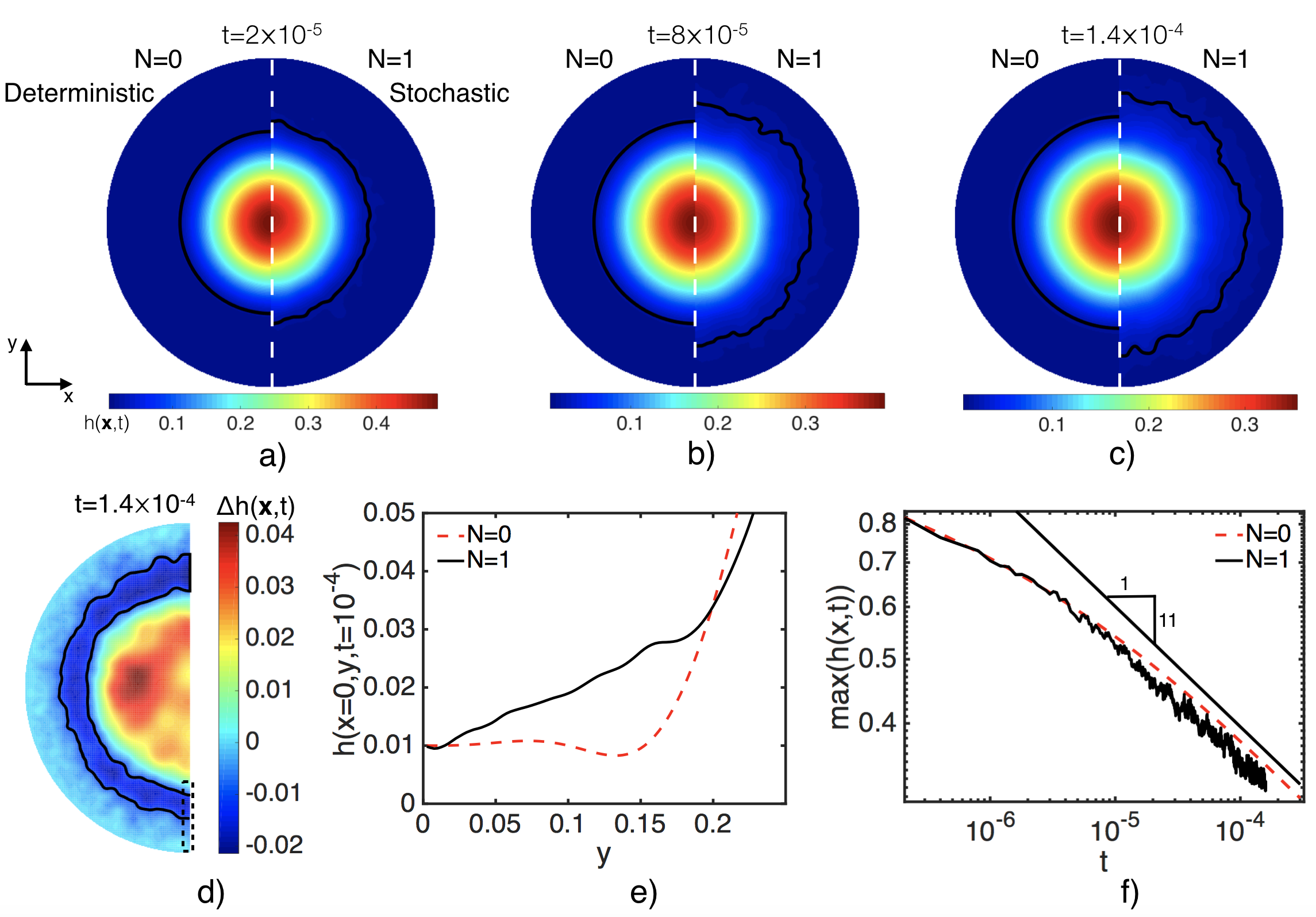}
\caption{Two-dimensional simulations of (\ref{eq:ndtf}), which represents the three dimensional dynamics height field $h(\mathbf{x},t)$ of the elastic blister. The simulations for $N=1$ is the average over 20 simulations. a-c) The spatiotemporal blister dynamics. Left panel: Deterministic simulation with $N=0$. Right panel: Inclusion of noise with $N=1$. The solid contour line shows the height $h(x,t)=0.03$ which is three times the height of the pre-wetted film layer $\epsilon$. d) To illustrate the difference between the deterministic and stochastic dynamics, we subtract the height profile for $N=1$ from $N=0$ i.e. $\Delta h(\mathbf{x},t)=h_{N=0}-\langle h \rangle_{N=1}$ at $t=1.4\times 10^{-4}$, showing that a precursor-like film is formed ahead of the bulk of the blister. e) We magnify the spreading front in the area denoted by the dashed lines in d), to show the local change in curvature and peeling of the elastic interface from the pre-wetted substrate. f) The spatiotemporal dynamics in a-c) suggest that the blister adopts to a self-similar profile. We extract $H(t)=\max(h(\mathbf{x,t}))$, located at the $\mathbf{x}$-position of the blister's center-of-mass. By plotting $H(t)$ in logarithmic axis shows that for $N=0$ the height evolves as $H(t)\sim t^{-1/11}$ as in (\ref{eq:det}), which is disturbed by the inclusion of noise $N=1$. 
\label{fig:fig2}}
\end{figure}

%\section{Stochastic elastohydrodynamics}
Previous studies on elastohydrodynamic flows have shown that the motion of an elastic sheet on a pre-wetted substrate is controlled by the curvature at its advancing front \cite{Lister2013PRL}. Thus, we suspect that any presence of fluctuations will cause local changes in the curvature, which will alter the sheet dynamics. To challenge this hypothesis, we considering the elastohydrodynamic analogue to Tanner's law i.e. a fluid filled blister with an elastic interface that deforms onto a pre-wetted substrate in the presence of fluctuations (see Fig. 1). The blister is characterized by its maximum height at its center of mass $\hat H(\hat t)$ and the spreading radius $\hat R(\hat t)$. The viscous fluid flow in the gap between the wall and the sheet is described by lubrication theory \cite{Batchelor,Oron:1997}  as we consider a blister with a lateral extent ($L$) that is much greater than its height $ \hat h= \hat h(\mathbf{\hat x}, \hat t)$ with $\nabla  \hat h\ll1$ i.e. we only consider long wavelength fluctuations. The elastic interface has a bending stiffness $B=Eb^3/(12(1-\nu))$ with $E$ the Young's modulus, $b$ the thickness of the elastic sheet and $\nu$ the poisson ratio. The pressure $ \hat p=\hat p(\mathbf{\hat x},\hat t)=B\nabla^4\hat h(\mathbf{\hat x},\hat t)$ is given by transverse force balance on the interface \cite{Landau1986}, which together with the inclusion of a stochastic stress term in the Navier Stokes equations result in a Focker-Planck like equation for the spatiotemporal viscous film \cite{grun2006},
\begin{equation}
\frac{\partial \hat h( \mathbf{\hat x}, \hat t)}{\partial \hat t}=\nabla\cdot(\frac{\hat h^3( \mathbf{\hat x}, \hat t)}{12\mu}\nabla \hat p(\mathbf{\hat x},\hat t))+\nabla\cdot(\Gamma \hat h(\mathbf{\hat x},\hat t)^{\frac{3}{2}}\hat\eta(\mathbf{\hat x}, \hat t))
\label{eq:tf}
\end{equation}
assuming no-slip at the two solid surfaces. The pre-factor multiplying the noise term is $\Gamma=\sqrt{\frac{2k_bT}{12\mu}},$ with the Bolzmann constant $k_b$, the temperature $T$ and the spatiotemporal Gaussian white noise $\hat \eta= \hat \eta( \mathbf{\hat x},\hat t)$ with zero mean and unit variance. Linear analysis of (\ref{eq:tf}) shows that it indeed recover the analytical results for the structure function \cite{grun2006}, as confirmed in simulations. We expect that the influence of fluctuations on the blister dynamics will increase in time as its curvature decreases, contrary to the coarsening of a capillary film where fluctuations are essential at short times \cite{Nesic2015,Diez2016a}. Thus, to accurately describe the blister dynamics we must capture the spatiotemporal interplay between its non-linear contribution in height and the multiplicative noise from thermal fluctuations in (2), of which an analytical solution is unfeasible. %We only consider long wavelength fluctuations and notice that linearization of the deterministic part of (\ref{eq:ndtf}) with $h=h_i +\epsilon\exp{kx-i\omega t}$ give
%\section{Dimensional analysis}
We make (\ref{eq:ndtf}) non-dimensional by scaling; $\mathbf{x}=\mathbf{\hat x}/L$ with the maximum blister extent $L$, the height ${h}={\hat h}/h_0$ with the maximum height of the blister at $\hat t=0$ i.e. $ h_0=\max({\hat h({0,\mathbf{\hat x}})})$ and scale time with $t= \hat t/\tau=t/\frac{12\mu L^6}{h_0^3B}$ giving the non-dimensional stochastic elastohydrodynamic thin film equation, 
\begin{equation}
\frac{\partial h(\mathbf{x},t)}{\partial t}=\nabla\cdot({h^3(\mathbf{x},t)}\nabla^5 h(\mathbf{x},t))+ N \times \nabla \cdot (h(\mathbf{x},t)^{\frac{3}{2}}\eta(\mathbf{x},t)).
\label{eq:ndtf}
\end{equation}
One non-dimensional number appears in the equation, $N=\sqrt{\frac{2k_B TL^2}{B h_0^2}}$ that gives the ration between the force generated by the thermal fluctuations and the elastic bending force from membrane deformation. To describe the deformation and spreading of the blister we deploy a scaling analysis and perform numerical simulations of (\ref{eq:ndtf}) in one and two dimensions by using the finite element method \cite{Amberg1999}. The deterministic parts of (\ref{eq:ndtf}) are solved implicitly \cite{Boyanova2012,carlson:2015}, while the noise-term $ \eta( \mathbf{ x},  t)$ is treated explicitly as in \cite{Nesic2015}. The numerical simulations are complemented by an initial condition for the blister height $ h(0, \mathbf{ x})=\epsilon+(1-\tanh(50(  x^2+  y^2)))$, where $ \epsilon$ is the pre-wetted layer and pinned boundary conditions imposed at the edge $\Gamma$ i.e.  $  h|_{\Gamma}( \mathbf{ x},  t)=  h|_{\Gamma}( \mathbf{ x}, t=0),\nabla^2   h|_{\Gamma} \cdot \mathbf{n}= p|_{\Gamma}=0$, with the boundary normal $\mathbf{n}$.
%\section{Numerical simulations and scaling analysis}
We investigate the influence of fluctuations on the elastic interface dynamics by solving (\ref{eq:ndtf}) with $N=0$ and $N=1$ in two-dimensional geometries, providing a three-dimensional description of the film. A direct comparison of the deterministic ($N=0$) and stochastic ($N=1$) simulation is shown in Fig. 2a-c, where we first notice that the major part of the bulk of the blister by eye appears evolve in a similar fashion. However, by subtracting the two height profiles (Fig. 2d) shows that the fluctuations affects the shape of the blister and ahead of its bulk a precursor-like film is formed as the elastic interface peel away from the pre-wetted substrate, see Fig. 2e. The blister dynamics appears to adopt a self-similar profile and the extraction of its height $H(t)$ in Fig. 2f shows that the deterministic simulation $N=0$ follows a power-law behavior $H(t)\sim t^{1/11}$ that is disturbed as $N>0$, see Fig. 2f.

To understand the blister dynamics in Fig. 2 we use a scaling analysis \cite{Lister2013PRL,Davidovitch:2005}, by starting to look at the limit where any fluctuation effects can be ignored as $N=0$. In this limit, the spreading of the blister is slow and has a nearly uniform pressure $\hat p=\hat p_0=B\nabla^4{\hat h({\mathbf{\hat x},\hat t})}\sim B \hat H(\hat t)/\hat R^4(\hat t)$ in its bulk \cite{Lister2013PRL}. By integrating $\hat p$ we get the blister shape by using axial symmetry, as also confirmed in the simulation, and boundary conditions; at the symmetry line $\nabla \hat h=\nabla^3\hat h=0$ and at the advancing front $\hat h=\nabla{\hat h}=\mathcal{O}(\epsilon)$ \cite{Lister2013PRL}. The blister shape combined with the local traveling-wave solution $\hat c$ gives the speed of the peeling tip $\frac{d\hat R}{d\hat t}=\hat c=\frac{Bh_0^{1/2}}{12\mu}({\frac{\kappa}{1.35}})^{5/2}$ \cite{Lister2013PRL} that is dictated by its local curvature $\kappa$. We scale $\hat h\sim \hat H(\hat t), \mathbf{\hat x}\sim \hat R(\hat t)$ together with the solution for $\hat c$ and impose mass conservation through a constant film volume $V=\int_0^{\hat R} \hat h(\mathbf{\hat x},\hat t) dV$ (in one dimension given by the constant area $A=\int_0^{\hat R} \hat h(\mathbf{\hat x},\hat t) dA$), which gives the self-similar spreading dynamics for a three dimensional blister
\begin{equation}
\hat R(\hat t)=1.81\left(\frac{Bh_0^{1/2}}{\mu}{V}^{5/2} \hat t\right)^{1/11},  \hat H(\hat t)=\frac{V}{\hat R^2(\hat t)}.
 \label{eq:det}
\end{equation}
In non-dimensional form this becomes $R(t)\approx t^{1/11}$ and $H(t)\approx t^{-2/11}$ and in one-dimensional geometries $R(t)\approx 7t^{2/17}$ and $H(t)\approx(1/7)\times t^{-2/17}$. This is the elastohydrodynamic analogue to Tanner's law for capillary spreading (2D: $R(t)\sim t^{1/10}$, 1D: $R(t)\sim t^{1/7}$), but contrast the classical solution with different exponents in time. Our self-similar solution is verified in the numerical solution shown in Fig. 2f with $N=0$. However, for $N>0$ the blister adopts a very different shape, which influences the spreading dynamics. 
\begin{figure}[h!]
\centering
\includegraphics[width=0.75\linewidth]{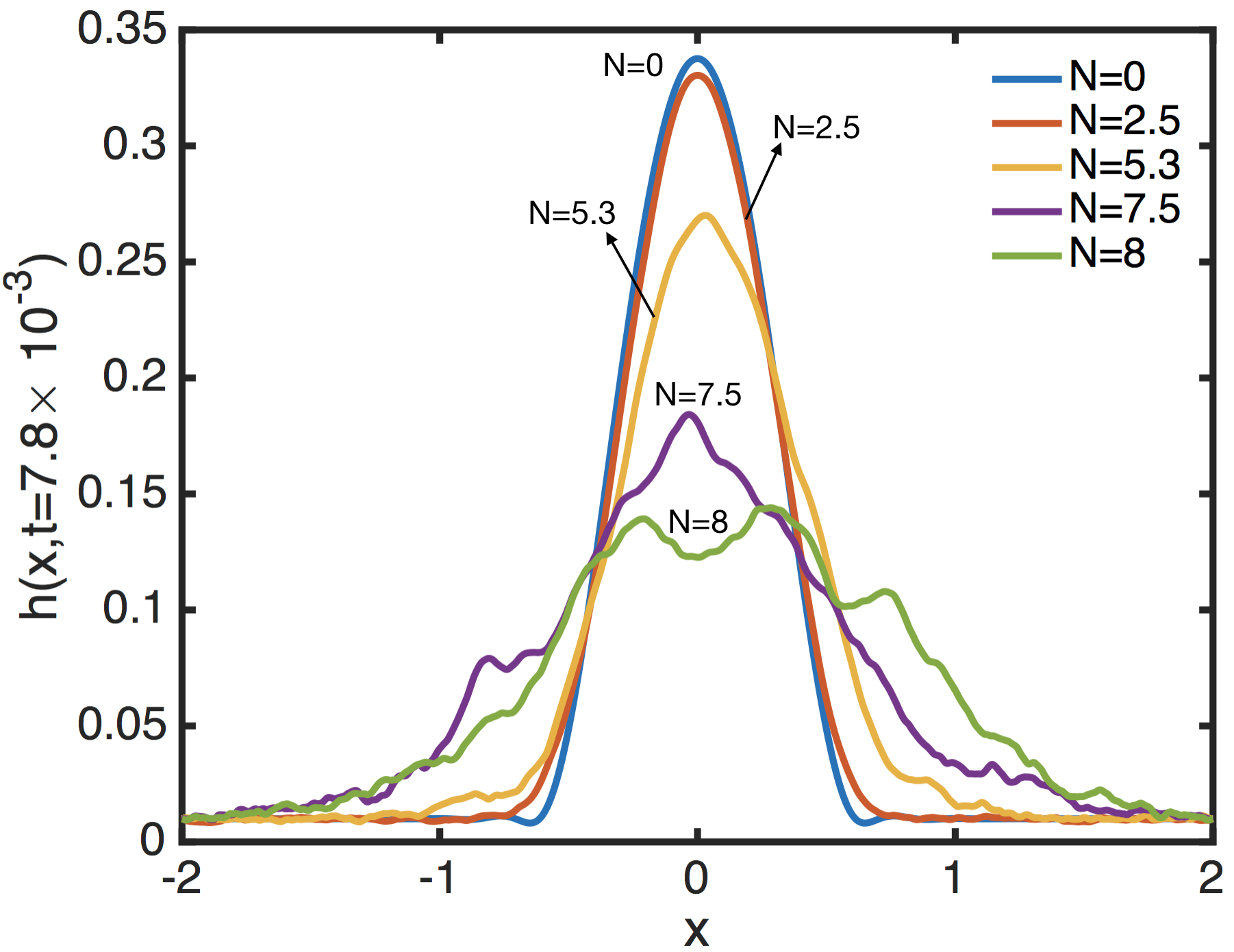}
\caption{Numerical simulations of (\ref{eq:ndtf}) in one dimension with $N\in [0, 2.5, 5.3, 7.5, 8]$ at time $t=7.8\times 10^{-3}$.
\label{fig:fig3}}
\end{figure}
\begin{figure}[h!]
\centering
\includegraphics[width=0.85\linewidth]{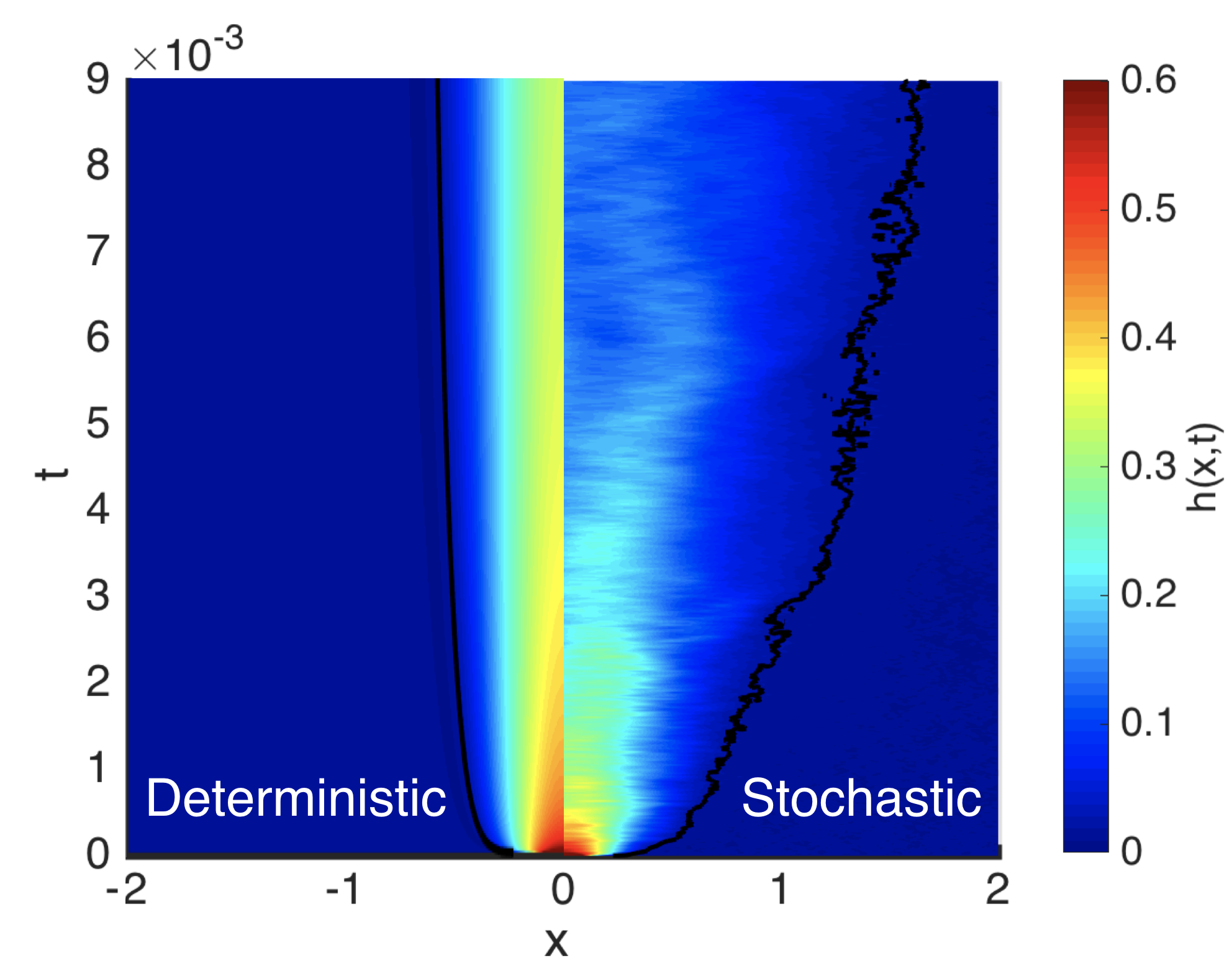}
\caption{Comparison of the spatiotemporal blister dynamics for $N=0$ (left panel) and $N=8$ (right panel). The solid line indicates the height $h(x,t)=0.03$ which is three times the pre-wetted film height $\epsilon$.
\label{fig:fig4}}
\end{figure}
\begin{figure}[h!]
\centering
\includegraphics[width=1.0\linewidth]{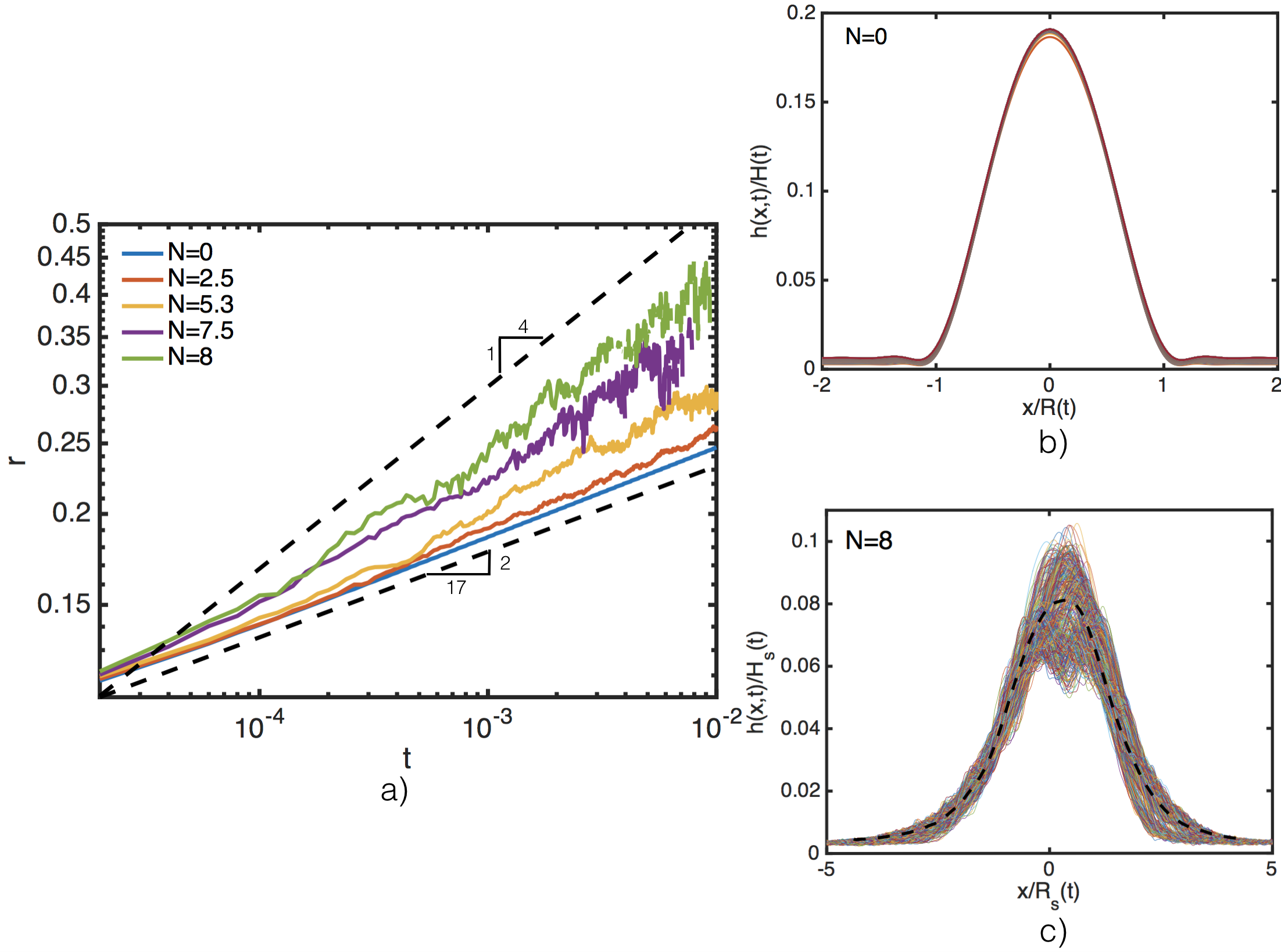}
\caption{a) The extracted spreading radius $r$ in logarithmic axis for $N\in [0, 2.5, 5.3, 7.5, 8]$ the dashed lines illustrate the spreading laws in (\ref{eq:det}-\ref{eq:ast}). b) The height profiles for the times $t=[1:2\times10^{-2}:10]\times 10^{-3}$ with $N=0$ are rescaled as $x/R(t)$ and $h/H(t)$ in (\ref{eq:det}). c) The height profiles for the times $t=[1:2\times10^{-2}:10]\times 10^{-3}$ with $N=8$ are rescaled as $x/R_s(t)$ and $h/H_s(t)$ in (\ref{eq:ast}), where the dashed line is the mean profile.
\label{fig:fig5}}
\end{figure}

Next, we examine the limit of a flaccid membrane i.e. $B\rightarrow0$ in (\ref{eq:tf}) to better understand how the thermal fluctuations influence the dynamics and look for its self-similar solution. The stochastic term scales as $\hat \eta(\mathbf{\hat x},\hat t)\sim\frac{1}{\sqrt{d\hat t} d\hat x}$ (in one dimensional geometries as $\hat \eta(\mathbf{\hat x},\hat t)\sim\frac{1}{\sqrt{d\hat t d\hat x\hat w}}$) \cite{Davidovitch:2005} in (\ref{eq:tf}), which combined with the blister's conservation of mass give, 
\begin{equation}
\hat R_s(\hat t)=(\frac{2k_bTV}{12\mu} \hat t)^{\frac{1}{6}},  \hat H_s(\hat t)=\frac{V}{ \hat R_s^2(\hat t)}.
\label{eq:ast}
\end{equation} 
In non-dimensional form this becomes $R_s(t)\approx (N t)^{1/6}$ and $H_s\approx R_s^2(t)\approx(N t)^{-1/3}$ and in one-dimensional geometries $R_s(t)\approx (N t)^{1/4}$ and $H_s(t)\approx (N t)^{-1/4}$, similar to the capillary films in \cite{Davidovitch:2005}. 

In order to characterize the influence of $N=\sqrt{\frac{2k_bTL^2}{Bh_0^2}}$ on the blister dynamics, we systematically vary $N$ in a one-dimensional geometry, see Fig. 3 and 4. It is clear that fluctuations can significantly speed-up the spreading dynamics and we observe a dramatic change in blister shapes. By tracking the spatiotemporal blister shape in Fig. 4 we identify two-distinct regimes, one for the elastohydrodynamic Tanner's law (\ref{eq:det}) and a stochastic spreading regime (\ref{eq:ast}).

We extract the spreading radius from the simulations in Fig. 3 by the averaged second moment to describe the advancing front
%\begin{equation}
$r=\Bigg \langle \int (x-X)h(x,t) dx \Bigg \rangle$
%\end{equation}
with $X=(\int x h(x,t) dx)/A$ \cite{Nesic2015}. The numerical simulations show that for $N=0$ the spreading recovers the deterministic self-similar profile in (\ref{eq:det}) and as $N\geq 5.3$ the advancing front adopts the stochastic dynamics predicted by (\ref{eq:ast}). These regimes are further illustrated by the collapse of blister shapes onto a universal curve by rescaling its numerical profiles with $x/R(t), h/H(t)$ for $N=0$ see Fig. 5b and $x/R_s(t), h/H_s(t)$ for $N=8$ see Fig. 5c.  

Our theoretical model is supported by simulations of (\ref{eq:ndtf}) in both one and two dimensions, but limited to fluctuations with a wavelength $\lambda \ll h/\nabla h$ which is asymptotically correct as $\nabla h\rightarrow 0$ for $\nabla t\rightarrow \infty$. Membrane dynamics at these length scales can also be influenced by a van der Waals pressure $\hat p_{vdW}=A/3\hat h^3$ with $A$ the Hamaker constant and for biological applications the adhesion of membrane embedded spring-like proteins $\hat p_s=K(1-\hat h/l_i)$ \cite{Leong}, with the effective spring stiffness $K$ and $l_i$ is the natural protein length, where we note that it is unlikely that these two effects would co-exist. To observe the stochastic regime, both of these additional terms in the pressure must be much smaller than the stochastic contribution in (\ref{eq:ndtf}). We use the scaling prediction in (\ref{eq:ast}) and compare the magnitude of these terms that could appear on the right hand side of (\ref{eq:ndtf}). The van der Waals adhesion scales as $\nabla\cdot(h^3\nabla p_{vdW})=\Gamma\nabla\cdot(h^{-1}\nabla h)\sim \Gamma R^{-2}$ with $\Gamma=\frac{AL^4}{Bh^4_0}$ a non-dimensional number for the ratio between van der Waals adhesion energy and the bending energy. Performing the same scaling analysis for the contribution from the spring pressure gives $\nabla\cdot(h^3\nabla p_{s})=S \nabla\cdot(h^{3}\nabla h)\sim S R^{-10}$ (in 1D $\sim  SR^{-6}$) with $S=KL^4/B$ a non-dimensional number for the ratio of the spring and the bending pressure. The contribution from bending the interface is $\nabla\cdot(h^3\nabla^5h)\sim R^{-14}$ (in 1D $\sim R^{-10}$) and the stochastic term scale as $\nabla\cdot({N h^{2/3}})\eta\sim N^2R^{-8}$ (in 1D $\sim N^2w^{1/2} R^{-5}$). 

We start by balancing the influence from the bending and the stochastic term, giving the condition $R\gg 1/N^{1/3}$ (in 1D $R\gg (w^{1/2}/N^2)^{1/5}$). Performing a similar analysis for the van der Waals contribution gives $R\ll (N^2/\Gamma)^{1/6}$ (in 1D $R\ll N/(\Gamma^2 w)^{1/6}$) and the spring pressure gives $R\gg (S/N^4)^{1/2}$ (in 1D $R\gg (S/N^4)^{1/2}$), these two additional conditions for $R$ needs to be realized in order to observe a stochastic spreading behavior. In the simulations we vary $N\in [0-8.5]$, which corresponds to a lipid membrane blister with a bending stiffness $B\in [10-100]k_bT$, at 300k with $k_bT=4\times 10^{-21}Nm$, a size $L=1\mu m$ and height $h_0=50nm$, which gives $N\in 3-9$ in accordance with the simulations presented in Fig. 2-5. A typical value for the Hamaker constant is $A\approx 10k_bT$  with $\Gamma\in[1.6 0.16]\times 10^5$. We assume a bio-inspired cell adhesion phenomenon \cite{carlson:2015} with a protein density $C\approx 100 (\mu m)^{-2}$ and stiffness $\gamma\approx 10^{-7}N/m$ giving $K=C\gamma=10^{7}N/m^3$ and $S\in [25-250]$.

We note that the condition for stochastic effects to dominate elastic bending $R> 1/N^{1/3}\approx 0.5$ (in 1D $R> (w^{1/2}/N^2)^{1/5}\approx 0.2$) and adhesive membrane embedded springs $R> (S/N^4)^{1/2}\approx0.2$ can be realized in an experiment and supports our numerical simulation in Fig. 2-5. The numerical simulations in Fig. 5 illustrates this transition in regimes as for $N=8$ and $N=7.5$ there is a clear change in slope for $R\approx 0.2$ in accordance with our analysis. Including van der Waals adhesion into the mathematical model creates another constraint on $R< (N^2/\Gamma)^{1/6}\approx 0.2$ (in 1D $R< N/(\Gamma^2 w)^{1/6}\approx 0.5$), which is challenging to realize together with the conditions from bending and spring pressure. {In biological membranes and synthetic lipid-membrane systems, adhesion is primarily modeled as adhesive springs \cite{Leong,Reister2008} where our stochastic elastohydrodynamic model gives an accurate description of the viscous flow in the membran gap that can help rationalize new regimes complementing those predicted by over-damped dynamics \cite{Reister2008}.}

%\section{Conclusions}
%Our stochastic elastohydrodynamic model couples viscous flow, elastic interface deformation and fluctuations from the environment, enabling a direct measure of the influence of these different contributions to the dynamics. 

The presented mathematical model presents a minimal and general theoretical skeleton for an accurate description of this class of stochastic elastohydrodynamic flows. By combining a scaling analysis and numerical simulations we find two spreading regimes; an elastohydrodynamic analogue to Tanner's law for capillary spreading and a stochastic dynamics, with a distinct transition between these two regimes also observed in the numerical simulations. Fluctuations alter the local curvature of the advancing front, leading to the peeling of the elastic interface leading to the formation of a precursor-like film advancing ahead of the bulk blister. This model is likely useful beyond the spreading of an elastic blisters as it sets the stage for the quantification of flow, elasticity, adhesion and stochastic effects in broader aspects of bio-inspired adhesion phenomena such as in vesicle interaction, protein organization in cell adhesion and growth of bio-films. \\

%\subsection{ACKNOWLEDGMENTS}
I would like to thank Dr. Minh Do-Quang for invaluable assistance in linking femLego with Boost. The financial support from the Norwegian Research Council -- NFR263056 and the Chair Total at ESPCI Paris are gratefully acknowledged. 
%\section{Methods}
%{The numerical simulation have been performed with the finite element code FemLego \cite{Amberg1999}. A solver based on Newtons method \cite{Boyanova2012} has been used to solve (\ref{eq:ndtf}), where the noise field has been created by a link to the open-source package from Boost.org. (\ref{eq:ndtf}) have been separated into three equations for the laplacian of $\Delta h$, the pressure $p$ and the height $h$, which have been discretized with linear elements. The deterministic parts of (2) are solved implicitly, while the noise term has been treated explicitly as in \cite{Nesic2015}. The spatial and temporal discretization has been varied between $\Delta x\in 0.025-0.06$ and $\Delta t\in[0.001-0.1]$. The mean data has been extracted from 20 two-dimensional runs in Fig. 1 and 40 one-dimensional runs in Fig. 3-5, where the computational time on a single CPU is about 4 weeks.}

\end{document}